\documentclass[journal]{IEEEtran}

\ifCLASSINFOpdf
\else
\fi

\usepackage{multirow}
\usepackage{mathtools}
\usepackage{amssymb}
\usepackage[noadjust]{cite}
\usepackage{threeparttable} 
\usepackage{siunitx}
\usepackage{colortbl}
\usepackage{xcolor}
\usepackage{etoolbox}
\usepackage{booktabs}
\definecolor{darkcyan}{rgb}{0.0, 0.55, 0.55}
\usepackage{tikz}
\usetikzlibrary{positioning, fit, arrows.meta, shapes}
\usetikzlibrary{calc}
\usepackage[T1]{fontenc}
\usepackage{enumerate}

\makeatletter
\patchcmd{\@classz}
  {\CT@row@color}
  {\oldCT@column@color}
  {}
  {}
\patchcmd{\@classz}
  {\CT@column@color}
  {\CT@row@color}
  {}
  {}
\patchcmd{\@classz}
  {\oldCT@column@color}
  {\CT@column@color}
  {}
  {}
\makeatother

\begin{document}

\title{Automated Polysomnography Analysis for Detection of Non-Apneic and Non-Hypopneic Arousals using Feature Engineering and a Bidirectional {LSTM} Network}

\author{Ali~Bahrami~Rad$^{*}$, Morteza~Zabihi, Zheng~Zhao, Moncef~Gabbouj, Aggelos~K.~Katsaggelos, and Simo~S\"{a}rkk\"{a}

\thanks{This work was supported in part by Business Finland Project 6015/31/2016 and in part by the Academy of Finland Project 313708. \textit{Asterisk indicates corresponding author.}}
\thanks{$^{*}$A.~B.~Rad is with the Department of Electrical Engineering and Automation, Aalto University, 02150 Espoo, Finland (e-mail: ali.bahrami.rad@aalto.fi).}
\thanks{M. Zabihi and M. Gabbouj are with the Department of Computing Sciences, Tampere University, 33014 Tampere, Finland.}
\thanks{Z. Zhao and S. S\"{a}rkk\"{a} are with the Department of Electrical Engineering and Automation, Aalto University, 02150 Espoo, Finland.}
\thanks{A. K. Katsaggelos is with the Department of Electrical Engineering and Computer Science, Northwestern University, Evanston, IL 60208, USA.}}

\markboth{}
{Shell \MakeLowercase{\textit{et al.}}: Bare Demo of IEEEtran.cls for Journals}

\maketitle

\begin{abstract}

\textit{Objective:} The aim of this study is to develop an automated classification algorithm for polysomnography (PSG) recordings to detect non-apneic and non-hypopneic arousals. Our particular focus is on detecting the respiratory effort-related arousals (RERAs) which are very subtle respiratory events that do not meet the criteria for apnea or hypopnea, and are more challenging to detect. \textit{Methods:} The proposed algorithm is based on a bidirectional long short-term memory (BiLSTM) classifier and 465 multi-domain features, extracted from multimodal clinical time series. The features consist of a set of physiology-inspired features ($n$ = 75), obtained by multiple steps of feature selection and expert analysis, and a set of physiology-agnostic features ($n$ = 390), derived from scattering transform. \textit{Results:} The proposed algorithm is validated on the 2018 PhysioNet challenge dataset. The overall performance in terms of the area under the precision-recall curve (AUPRC) is 0.50 on the hidden test dataset. This result is tied for the second-best score during the follow-up and official phases of the 2018 PhysioNet challenge. \textit{Conclusions:} The results demonstrate that it is possible to automatically detect subtle non-apneic/non-hypopneic arousal events from PSG recordings. \textit{Significance:} Automatic detection of subtle respiratory events such as RERAs together with other non-apneic/non-hypopneic arousals will allow detailed annotations of large PSG databases. This contributes to a better retrospective analysis of sleep data, which may also improve the quality of treatment.
 
\end{abstract}


\IEEEpeerreviewmaketitle

\section{Introduction}

\IEEEPARstart{M}{edical} studies show a bidirectional relationship between sleep and health, and consequently, sleep disorders may have a negative effect on patients' health, mood, and quality of life~\cite{avidan2011handbook}. There are about 90 different sleep disorders classified under the main categories of insomnia, sleep-related breathing disorders, sleep-related movement disorders, hypersomnias of central origin, parasomnias, and circadian rhythm sleep disorders~\cite{pandi2016epidemiology2}. In this study, we pay special attention to sleep arousals induced by sleep-related breathing disorders. However, sleep arousals which are identified by transitions from deeper sleep states to lighter ones can also occur either spontaneously or in association with other sleep disorders and/or environmental stimuli.

Sleep arousals are characterized by sudden shifts in electroencephalography (EEG) frequency~\cite{bonnet1992asda}. However, depending on the type of sleep disorders, arousals may be manifested on other biosignals too. For example, sleep-related breathing disorders, which are characterized by respiratory or ventilatory disturbance during sleep~\cite{kirsch2013sleep}, lead to arousals detectable from biosignals such as airflow, respiratory effort signals (chest and abdominal), and arterial oxygen saturation (SaO$_2$) along with EEG. Furthermore, bruxism, defined as unconscious clenching, grinding, or bracing of the teeth during sleep~\cite{pandi2016epidemiology}, is a type of sleep-related movement disorder which leads to arousals observable from chin EMG and EEG~\cite{salas2017sleep}. Therefore, analysis of the patterns of the aforementioned clinical time series together with other biosignals such as electrooculography (EOG) and electrocardiography (ECG), which are recorded during a typical polysomnography (PSG) test, provide important information for sleep arousal detection. 

Despite the recent attempts to automate PSG-based sleep analysis~\cite{sun2017large,biswal2018expert,stephansen2018neural,8631195,8502139,cooray2019detection,malafeev2018automatic}, arousal detection is still done manually by expert sleep technologists. Typical contemporary PSG datasets can consist of hundreds to thousands of cases, and each case contains more than a dozen clinical time series with about eight-hour long. Manual analysis of such datasets is a labor-intensive and time-consuming process, which highly depends on the sleep technologist's experience and skill~\cite{penzel2007digital}, and consequently limits the PSG-based sleep-related studies. Our aim is to develop a machine learning algorithm to automatically detect arousal events in PSG recordings. We use the same objective as appointed by the PhysioNet/Computing in Cardiology (CinC) Challenge 2018~\cite{CinC2018,ghassemi2018you}. According to the PhysioNet challenge rules, the target arousals are those which are neither apneic nor hypopneic. This excludes all apnea types including obstructive, central, and mixed events~\cite{rosenberg2014american} as well as hypopnea from our analysis. 

Our particular focus is on detecting the respiratory effort-related arousals (RERAs) which account for 99.6\% of all target arousals available in the PhysioNet training dataset~\cite{ghassemi2018you}. RERA is a sequence of breaths lasting at least 10 seconds, characterized by extended inspiratory phase, paradoxical movement of the chest and abdomen, and/or flattening of inspiratory airflow that leads to an arousal from sleep~\cite{chokroverty2017oxford,berry2012aasm}. RERAs are very subtle respiratory events which do not meet criteria for apnea or hypopnea and are more challenging to detect~\cite{cracowski2001characterization}. Despite its subtle nature and moderate manifestation on biosignals, RERAs can cause fatigue and daytime sleepiness~\cite{guilleminault1993cause}, not to mention an excessive number of RERAs is also associated with raised blood hypertension~\cite{guilleminault1996upper} and car accidents~\cite{masa2000habitually}. Aside from RERAs, the remaining 0.4\% target arousals of this study consist of other types of sleep-related breathing disorders, sleep-related movement disorders, environmental stimuli, and spontaneous arousals. 

The current study is a continuation of our prior work~\cite{zabihi2018automatic} in the sense that it is developed for the follow-up phase of the 2018 PhysioNet challenge, and then assessed on the same dataset with the same evaluation criteria. However, it is a thoroughly independent body of research by virtue of the following facts. First, in our prior work, we proposed an automatic feature learning procedure based on a 2D convolutional neural network (CNN)~\cite{lecun2015deep} and state distance representation~\cite{hatami2018classification} of clinical time series. However, in the current study, we extract hand-engineered features from various time series based on the combination of expert knowledge and feature selection techniques. Second, in our previous study, we used a limited number of PSG channels (only 3 biosignals), but here we use all available PSG data (13 biosignals). Third, the development of the previous algorithm involved minimum/no physiological knowledge; the currently proposed method is developed based on prior knowledge of the physiological process during arousals. Fourth, we also extract an alternative semi-automatic set of features using state-of-the-art scattering transform~\cite{mallat2012group} and investigate ways to increase its performance for PSG classification. Fifth, we utilize a recurrent neural network (RNN) based on long short-term memory (LSTM) units~\cite{hochreiter1997long} for sequence modeling of sleep microstructures and transient events. The developed software is available in the PhysioNet system and will be released under an open-source license, according to the PhysioNet timeline.

\section{Dataset and Evaluation Criteria}\label{sec:dataset}

We use the same dataset and scoring mechanism as provided by the 2018 PhysioNet/CinC challenge. The dataset comprises 1983 cases of in-laboratory PSG recordings. The data were recorded by Massachusetts General Hospital's (MGH) Sleep Lab in the Sleep Division together with the Computational Clinical Neurophysiology Laboratory, and the Clinical Data Animation Center according to the American Academy of Sleep Medicine (AASM) practice standards~\cite{ghassemi2018you}. The recordings consist of 13 biosignals as follow: 
\begin{itemize}
\item six EEG channels for recording cortical activity of three brain regions, based on the International 10-20 System:
\begin{itemize}
\item frontal: F3-M2 and F4-M1 (PSG channels 1, 2),
\item central: C3-M2 and C4-M1 (PSG channels 3, 4), 
\item occipital: O1-M2 and O2-M1 (PSG channels 5, 6); 
\end{itemize}
\item the left side EOG for recording eye movements (PSG channel 7);
\item chin EMG for measuring chin muscle activity (PSG channel 8); 
\item two respiratory  effort  signals for recording thoracoabdominal movements:
\begin{itemize}
\item abdominal (PSG channel 9), 
\item chest (PSG channel 10); 
\end{itemize}
\item respiratory airflow (PSG channel 11);
\item arterial oxygen saturation (SaO$_2$) (PSG channel 12);
\item ECG for measuring heart activity (PSG channel 13).
\end{itemize}
All biosignals except SaO$_2$ were sampled at 200~Hz. The SaO$_2$  was upsampled to 200~Hz for convenience.

All recordings were annotated according to AASM standard by seven clinical experts, but one expert was used for each recording. The recordings were scored for sleep stages and then annotated into three classes: non-target arousal, target arousal, and non-arousal events. The non-target arousals are those regions in PSG recordings with apneic or hypopneic arousals, and the target arousals are the regions which meet either of the following conditions: 
\begin{enumerate}[(i)]
\item 2 seconds before the onset of RERA to 10 seconds after its ending;
\item 2 seconds before the onset of non-RERA, non-apneic, and non-hypopneic arousal to 2 seconds after its ending.
\end{enumerate}

As it was stated earlier, 99.6\% of target arousals in the training dataset are related to RERAs. The remaining 0.4\% are distributed among arousals related to snoring, partial airway obstruction, Cheyne-Stokes breathing, hypoventilation, bruxism, periodic leg movement, noise, and spontaneous.

The dataset is divided into two disjoint subsets of training ($n$ = 994 subjects) and testing ($n$ = 989). The labels of the testing dataset are hidden and are reserved to be used by PhysioNet challenge organizers to evaluate the performance of the submitted algorithms. The performance is assessed using the area under the precision-recall curve (AUPRC) for binary classification between target arousal and non-arousal regions. The non-target arousal regions are not considered for evaluation. More information on the evaluation criteria is available in~\cite{CinC2018} and~\cite{ghassemi2018you}. In addition to AUPRC which is the primary evaluation criterion, we calculated the area under the receiver operating characteristic curve (AUROC) as a secondary evaluation criterion.

\section{Feature Engineering}

After preprocessing of PSG recordings as described in Section~\ref{preprocessing}, we extract 465 features from each 5-second analysis window. The features are categorized into two groups: physiology informed and physiology agnostic features. The physiology informed features are extracted based on our physiological knowledge of sleep arousal and its manifestations on biosignals. However, this set of features are not solely based on physiology, but instead, we extract a large number of hand-engineered features based on our prior knowledge of sleep arousals, and then during multiple steps of feature selection and expert judgments, remove the irrelevant and/or redundant ones (see Section~\ref{PIF}). On the other hand, the physiology agnostic features are entirely derived based on our knowledge of signal processing and machine learning without any physiological consideration (see Section~\ref{sec:ST}).

\subsection{Preprocessing and Data Preparation}\label{preprocessing}

The 60~Hz powerline artifact is removed using a band-stop filter. Moreover, an inspection of the spectral content of biosignals indicates the presence of an extra 80~Hz artifact in some recordings. This might be related to the second harmonic of the powerline artifact (120~Hz) which due to the aliasing effect presents itself as an 80~Hz false frequency component. The 80~Hz artifact is filtered out as well. Then, the high-amplitude muscle-generated artifacts due to body movements are removed by simple thresholding: if the instantaneous magnitude of the biosignal exceeds 8 times the interquartile range of its amplitude, it is replaced by zero value. Furthermore, the dynamic range of the signal amplitude is normalized by dividing the instantaneous amplitude by 8 times the interquartile range. The last two steps (i.e., high-amplitude artifact removal and dynamic range normalization) are applied to all biosignals except SaO$_2$ and ECG. Finally, each PSG recording is segmented into 5-second nonoverlapping triangular windows. From now on, all the analyses are done on these 5-second windows.

\subsection{Physiology Informed Features}\label{PIF}

In the initial phase of this study, we extracted more than 900 features from all biosignals. The extracted features were from various domains such as time, frequency (or spectral), time-frequency, and phase space. The number of features is then reduced through multiple steps of feature selection methods and expert judgment. In the first step, 250 features are removed after a feature ranking procedure using a random forest classifier similarly to~\cite{zabihi2017detection} and~\cite{isasiecg}. Then, the features derived by nonlinear analysis of biosignals in the reconstructed phase space~\cite{takens1981detecting} are removed to speed up the feature extraction process. Although these features contribute to a better classification result by $\sim$0.02 points in terms of AUPRC, we remove them from our analysis due to the run-time constraint applied by the PhysioNet challenge organizers. In the next step, all time-frequency features, obtained from the ordinary discrete wavelet transform (DWT), are removed and replaced by features derived from the scattering transform~\cite{mallat2012group}. The problem with DWT is that it is covariant to translation and one needs to extract the ad hoc translation invariant features similarly to~\cite{rad2017ecg}. Since the calculation of scattering transform features involves no physiological knowledge, we treat them as physiology agnostic features, discussed separately in Section~\ref{sec:ST}. In the last step, we applied our proposed heuristic feature selection method (see Section~\ref{sec:classification}) to the remaining 500 features to derive the final set of 75 physiology informed features.

In the following, we describe these 75 features which can be further categorized into two subgroups: respiratory-related and non-respiratory-related features. The respiratory-related features, described in Section~\ref{sec:RRF}, are extracted from biosignals related to the respiratory process such as abdominal, chest, airflow, and SaO$_2$. The non-respiratory-related features, described in Section~\ref{sec:NRRF}, are extracted from EEGs, EOG, chin EMG, and ECG.

\subsubsection{Respiratory-related features}\label{sec:RRF}

Monitoring respiratory activity using relevant biosignals such as airflow, abdominal and chest, as well as oxygen saturation (SaO$_2$) reveals abnormalities and/or complications related to breathing~\cite{thorpy2010parasomnias}. For example, SaO$_2$ indicates changes in blood oxygen level which is an important marker for the detection of sleep apnea or other respiratory problems. The respiratory-related biosignals also capture information about snoring, respiratory rate, airway obstruction, and the strength of inhalation and expiration~\cite{pagel2014primary}. For instance, the morphology and movement patterns of the chest and abdomen (e.g., biphasic, paradoxical, and in-phase) and/or the shape of the airflow signal (flatten vs. normal) are important indicators for detection of RERAs~\cite{masa2003assessment,goldman1995asynchronous}. Furthermore, snoring can be derived from the high-frequency periodic oscillation of airflow~\cite{krahn2010atlas}, or it might even appear as an artifact on the non-respiratory-related chin EMG biosignal~\cite{chokroverty2017oxford}.  

In the following, we describe the selected 41 respiratory-related features, among them, there are 13 cross-channel and 28 isolated-channel features. 

\begin{enumerate}[(i)]
  \item Thirteen cross-channel features are extracted from the abdominal, chest, and airflow signals (PSG channels 9-11) using correlation analysis, hypothesis testing, and multichannel signal decomposition. The first six features are the Pearson correlation coefficients and the p-values for testing the hypothesis that there is no relationship between each pair of signals (null hypothesis). The next seven cross-channel features are extracted after factorization of the the matrix $\boldsymbol{X}$ formed by these signals each of length $N$ 
\begin{equation}
\boldsymbol{X}= 
\begin{bmatrix}
   x^9_1 & x^9_2 & \cdots & x^9_{N}\\
  x^{10}_1 & x^{10}_2 & \cdots & x^{10}_{N}\\
    x^{11}_1 & x^{11}_2 & \cdots & x^{11}_{N}\\
\end{bmatrix}.
 \end{equation}
We consider the singular value decomposition (SVD) of $\boldsymbol{X}$~\cite{christensen2016introduction}, that is,
\begin{equation}
\boldsymbol{X}= \boldsymbol{U} \begin{bmatrix}\boldsymbol{\Sigma} & \boldsymbol{0}\end{bmatrix} \boldsymbol{V}^\top,
 \end{equation}
where $\boldsymbol{U}$ and $\boldsymbol{V}$ are $3 \times 3$ and $N \times N$ orthogonal matrices, respectively, and $\begin{bmatrix}\boldsymbol{\Sigma}& \boldsymbol{0}\end{bmatrix}$ is a $3 \times N$ block matrix in which $\boldsymbol{\Sigma}$ is a $3 \times 3$ diagonal matrix with singular values $\sigma_1$, $\sigma_2$, and $\sigma_3$ in the diagonal, that is,   
\begin{equation}
\boldsymbol{\Sigma} = 
\begin{bmatrix}
   \sigma_1 & 0 & 0 \\
  0 & \sigma_2 & 0 \\
    0 & 0 & \sigma_3 \\
    \end{bmatrix},
  \end{equation}
and $\boldsymbol{0}$ is a $3 \times (N-3)$ zero matrix. By convention $\boldsymbol{U}$, $\boldsymbol{V}$, and $\boldsymbol{\Sigma}$ are organized such that $\sigma_1 \geq \sigma_2 \geq \sigma_3 \geq 0$. The singular values $\sigma_1$, $\sigma_2$, $\sigma_3$, their arithmetic and geometric means, their standard deviation (STD), and the ratio of $\sigma_1/\sigma_3$ are then the next seven cross-channel features.

\begin{figure*}[t!]
	\centering
	\includegraphics[width=0.7\linewidth]{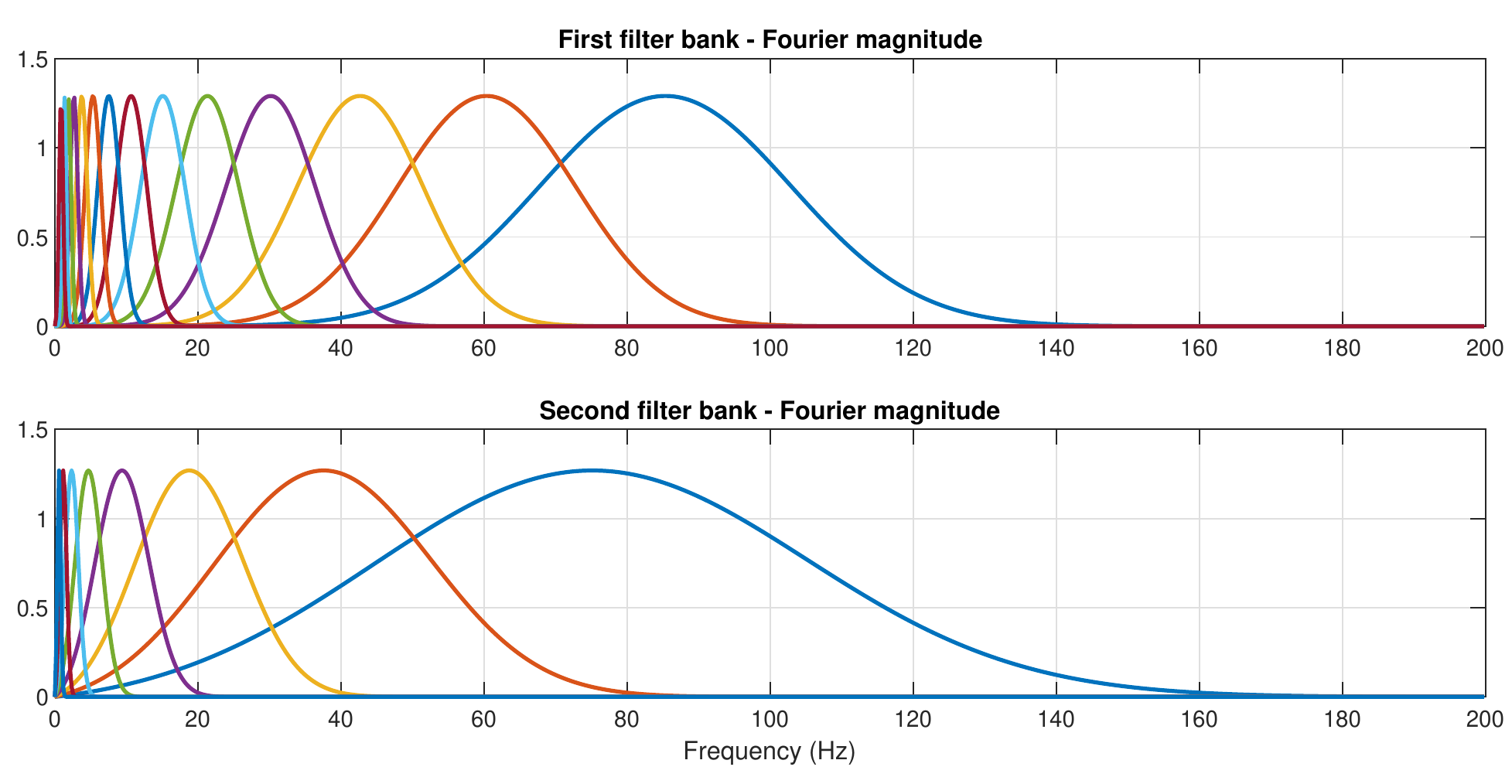}
	{\caption{Magnitudes of the frequency spectra of the wavelets in the two filter banks. In the first filter bank $Q=2$, $J = 13$, and $P=1$, thus the number of wavelets is 14 ($=J+P$). In the second filter bank $J=8$ and since $Q = 1$, $P=0$. Thus, the number of wavelets is 8 ($=J+P$).}\label{fig:filterbanks}}
\end{figure*}

\item Six features are extracted from the abdominal signal. The first two features are the STD and the root mean square (RMS) values of the signal. Then the signal is modeled as an order 10 autoregressive (AR) process, that is, 
\begin{equation}\label{ARmodel}
x_n=-\sum_{k=1}^{10} a_k x_{n-k}+v_n,
\end{equation}
where $x_n$ and $v_n$ are the $n$-th samples of the signal and input white noise, respectively, and $a_1,a_2,\cdots,a_{10}$ are the parameters of the model, estimated by Burg's method~\cite{Burg-68}. The third feature is the ninth parameter ($a_9$) of the above-mentioned AR model. It is worth mentioning that there are various approaches for choosing a good value for the order of the AR model such as minimizing either the Akaike or the Bayesian information
criteria~\cite{Akaike:1974,schwarz1978estimating}. However, in this work, our purpose is not to design an optimum model for signal representation, but we are merely looking for those parameters (i.e., features) that are informative enough to be used for discrimination between arousal and non-arousal classes. Therefore, instead of being preoccupied with the optimum model selection, we choose a model order with a moderate value (e.g., 10) and during a feature selection procedure, choose the discriminative parameters. Furthermore, the respiratory-related abdominal spectrum (i.e., low-frequency interval of the abdominal spectrum) is divided into the following five frequency bands: 0.01-0.4~Hz, 0.4-0.75~Hz, 0.75-1.2~Hz, 1.2-1.6~Hz, and 1.6-3~Hz. The signal power in the frequency band between $f_1$-$f_2$~Hz, $\mathcal{P}{(f_1,f_2)}$, is estimated by the area under the power spectral density curve, $\hat{P}(f)$, that is, 
\begin{equation}\label{PSD}
\mathcal{P}{(f_1,f_2)} = \int_{f_1}^{f_2} \hat{P}(f)\: \mathrm{d}f,
\end{equation}
where $\hat{P}(f)$ is estimated using Burg's AR model with an empirically derived order 30. The last three features are $\mathcal{P}{(0.01,0.4)}$, $\mathcal{P}{(0.4,0.75)}$, and the ratio of $\mathcal{P}{(0.75,1.2)}/\mathcal{P}{(1.2,1.6)}$.     

\item Five features are extracted from the chest signal, namely, RMS, STD, skewness, $\mathcal{P}{(0.01,0.4)}$, and $\mathcal{P}{(0.75,1.2)}/\mathcal{P}{(1.2,1.6)}$.   

\item Twelve features are extracted from airflow. We extracted RMS and skewness of the signal along with its power in five frequency bands: $\mathcal{P}{(0.01,0.4)}$, $\mathcal{P}{(0.4,0.75)}$, $\mathcal{P}{(0.75,1.2)}$, $\mathcal{P}{(1.2,1.6)}$, and $\mathcal{P}{(1.6,3)}$. Moreover, the next four features are the nonlinear combinations of these features as follow: $\mathcal{P}{(0.4,0.75)}\times\mathcal{P}{(1.2,1.6)}$, $\mathcal{P}{(0.75,1.2)}\times\mathcal{P}{(1.2,1.6)}$,  
$\mathcal{P}{(0.75,1.2)}/\mathcal{P}{(1.2,1.6)}$, and 
$\mathcal{P}{(0.01,0.4)}/(\mathcal{P}{(0.75,1.2)}+\mathcal{P}{(1.6,3)})$. The last feature is $(\mathrm{STD}(\ddot{x})\times\mathrm{STD}(\dot{x}))/\mathrm{STD}(x)$, in which $x$, $\dot{x}$, and $\ddot{x}$ are the airflow signal and its first and second forward differences, respectively, that is,   
\begin{equation}\label{eq:diff}
  \begin{aligned}
& \dot{x}_n = {x}_{n+1}-{x}_n,\\ 
& \ddot{x}_n = {x}_{n+2}-2{x}_{n+1}+{x}_n.
\end{aligned}
\end{equation}
   
\item We extract 5 features from SaO$_2$, namely, mean, STD, RMS, and mean frequency of the power spectrum of the signal, as well as STD of the signal first difference. 

\end{enumerate}

\begin{figure*}[t!]
	\centering
	\includegraphics[width=0.7\linewidth]{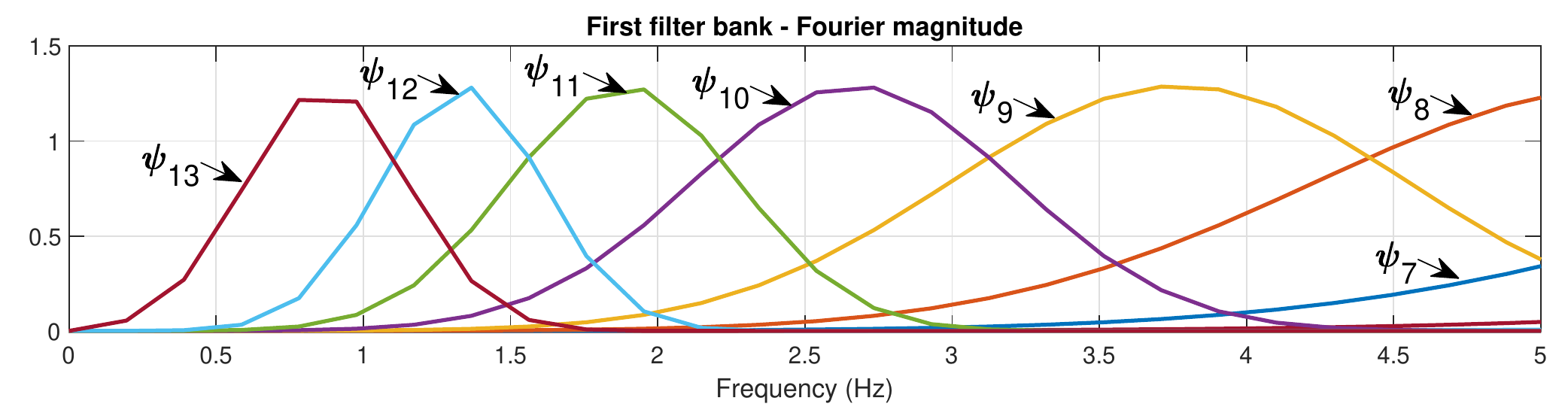}
	{\caption{Magnitudes of the frequency spectra of the wavelets in the first filter bank of Fig.~\ref{fig:filterbanks} are shown in the interval from 0 to 5~Hz. $\psi_{13}$ has the same bandwidth as $\psi_{12}$, but the bandwidths of $\psi_{12}$ to $\psi_{0}$ are increasing exponentially.}\label{fig:filterbanks2}}
\end{figure*}

\subsubsection{Non-respiratory-related features}\label{sec:NRRF}

AASM guidelines define arousal as an abrupt shift in EEG frequency including alpha (8-13~Hz), theta (4-8~Hz), and frequencies above 16~Hz lasting at least 3 seconds, and is preceded by at least 10 seconds of stable sleep~\cite{berry2018aasm}. Moreover, during a rapid eye movement (REM) stage, this EEG frequency shift needs to be accompanied by concurrent increases in submental (chin) EMG amplitude, to be recognized as arousal. On the other hand, non-rapid eye movement (NREM) arousals may occur without the aforementioned increase in chin EMG.  

We extract various features from different EEG frequency bands including delta (0.1-4~Hz), theta (4-8~Hz), alpha (8-13~Hz), sigma (13-16~Hz), and beta (16-25~Hz). Moreover, EOG- and EMG-based features are extracted to differentiate between EEG arousals during REM and NREM, and ECG-based features are extracted to provide complementary information about sleep-related breathing disorders as well as autonomic arousals~\cite{zaiwalla2017polysomnography}. In the following, the selected 34 non-respiratory-related features are described.

\begin{enumerate}[(i)]
  \item Seven features are extracted from each frontal EEG and EOG (PSG channels 1, 2, 7) as follow: RMS, STD, skewness, and kurtosis of biosignals, together with $a_3$ and $a_5$ parameters of AR  model in~(\ref{ARmodel}) and $\mathcal{P}{(0.1,4)}$ in~(\ref{PSD}).
  \item RMS and $a_3$ in~(\ref{ARmodel}) are calculated for each central and occipital EEG (PSG channels 3-6).

  \item Three features are extracted from chin EMG as follow: RMS and kurtosis of the signal, in addition to $\mathcal{P}{(0.1,15)}/(\mathcal{P}{(30,45)}+\mathcal{P}{(70,100)})$.    
  \item The following two features are extracted from ECG signal: $\mathcal{P}{(7.5,12)}/\mathcal{P}{(12,16)}$ and  $\mathcal{P}{(12,16)}/(\mathcal{P}{(7.5,12)}+\mathcal{P}{(16,25)})$.
\end{enumerate}

\subsection{Physiology Agnostic Features}\label{sec:ST}

\begin{figure*}[t!]
	\centering
	\includegraphics[width=0.8\linewidth]{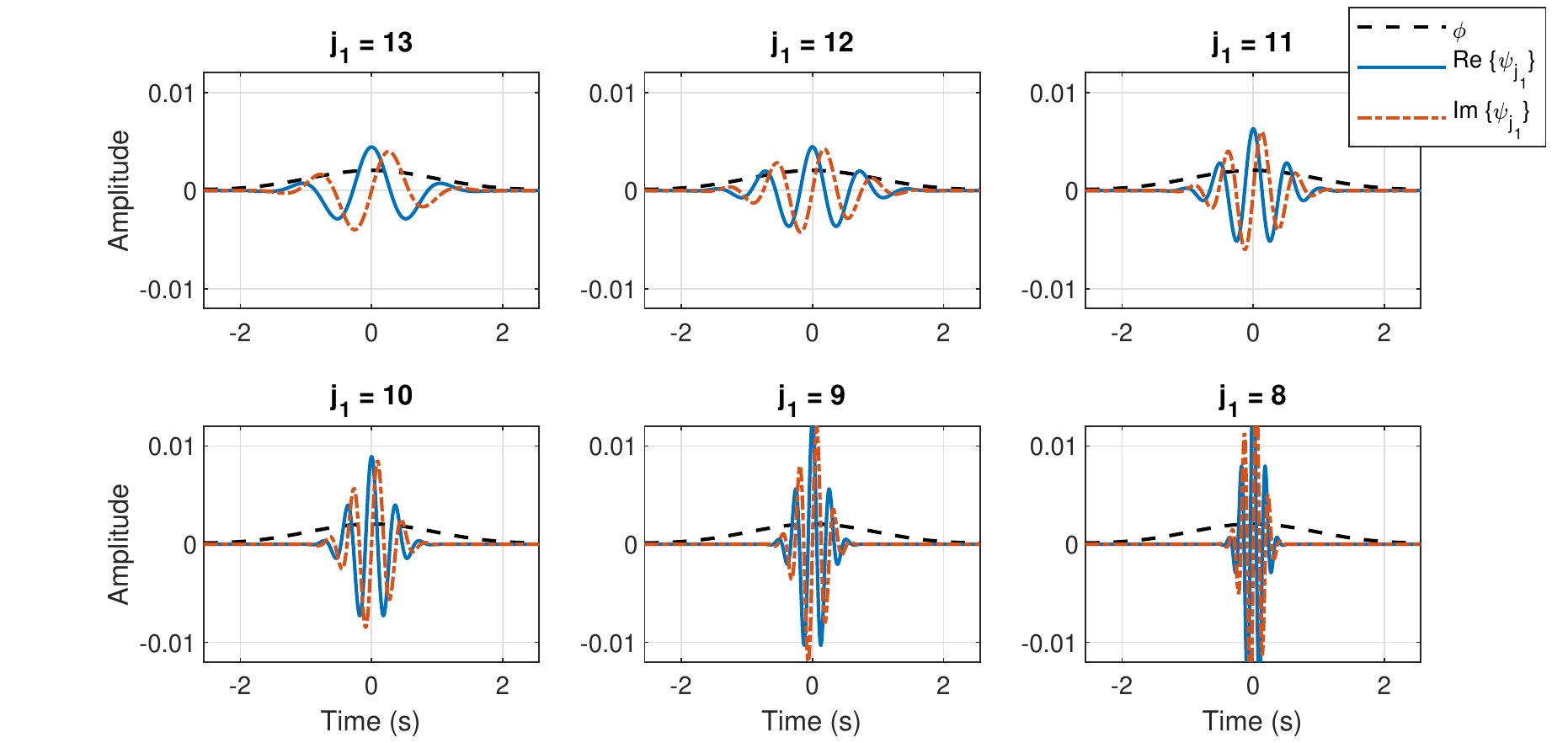}
	{\caption{Time-domain representation of the complex wavelets corresponding to the analytical filters in Fig.~\ref{fig:filterbanks2}. Re$\{\psi{j_1}\}$ and Im$\{\psi{j_1}\}$ are the real and imaginary parts of the complex wavelet functions corresponding to analytical filters shown in Fig.~\ref{fig:filterbanks2}. $\phi$ is the approximation function whose corresponding low-pass filter is not shown in Fig.~\ref{fig:filterbanks2}. }\label{fig:filterbanks3}}
\end{figure*}

One of the challenges in classification is handling a substantial amount of intra-class variability which is not helpful for discrimination between different classes. Removing or minimizing this irrelevant information and preserving useful inter-class variabilities may significantly increase the classifier's performance. The scattering transform proposed by Mallat~\cite{mallat2012group} is a systematic approach to address this problem by building locally invariant, stable, and informative representations while preserving the signal norm and most of the inter-class variabilities.

The scattering transform is a deep representation which mimics a CNN in the sense that it propagates the input signal across a sequence of linear filters followed by pooling and nonlinearities~\cite{mallat2016understanding}. However, contrary to a CNN in which the filters have adaptive weights obtained through a gradient-based learning strategy and error back-propagation~\cite{rumelhart1988learning}, the scattering transform is derived by cascading predefined filters, namely wavelets. To be more precise, the scattering transform is a deep signal representation, derived by cascading wavelet transform moduli followed by an averaging operator (i.e., low-pass filtering)~\cite{mallat2012group}. The logic behind this new transformation is to derive a translation invariant representation of the original signal which is also stable to small deformations like time warping.  

In this work, we design a 2-layer scattering network with corresponding filter banks illustrated in Fig.~\ref{fig:filterbanks}. We use Gabor wavelets (i.e., approximately analytic wavelets constructed by frequency modulation of Gaussian windows)~\cite{Mallat:2008} whose central frequencies of the mother wavelets in the first and second filter banks are calculated as follows
\begin{eqnarray}
         \omega_0^{\langle 1 \rangle} =  \left(\frac{1+2^{-1/Q_1}}{2}\right)\times \frac{f_s}{2} = 85.35~\textnormal{Hz}, \\
         \omega_0^{\langle 2 \rangle} =  \left(\frac{1+2^{-1/Q_2}}{2}\right)\times \frac{f_s}{2} = 75.00~\textnormal{Hz}.
\end{eqnarray}
Here, quality factors $Q_1 = 2$ and $Q_2=1$ are the number of wavelets per octave for the first and second filter banks, and $f_s = 200$~Hz is the sampling frequency. We design this scattering network such that the resulting representation is invariant to 5-second translation which leads to $J_1 = 13$ and $J_2 = 8$ wavelet scales in the first and second filter banks. Other wavelets $\psi_{j_k}^{\langle k \rangle}(t)$ in the filter banks are derived by dilating the mother wavelets $\psi_{0}^{\langle k \rangle}(t)$ by a factor of $2^{1/Q_k}$  
\begin{equation}
\psi_{j_k}^{\langle k \rangle}(t) = 2^{-{j_k}/{Q_k}}\psi_{0}^{\langle k \rangle}(2^{-{j_k}/{Q_k}}t),
\end{equation}
where $k \in \{1, 2\}$ indicates the layer index in the scattering network and $j_k \in \{0,1,2,\cdots,J_k-1\}$ indicates the scale index. In the Fourier domain, these filter banks can be represented by
\begin{equation}
\hat\psi_{j_k}^{\langle k \rangle}(\omega) = \hat\psi_{0}^{\langle k \rangle}(2^{{j_k}/{Q_k}}\omega),
\end{equation}
whose magnitudes are demonstrated in Fig.~\ref{fig:filterbanks}. If the central frequency of $\hat\psi_{0}^{\langle k \rangle}(\cdot)$ is $\omega_{0}^{\langle k \rangle}$, then the central frequency of $\hat\psi_{j_k}^{\langle k \rangle}(\cdot)$ is $2^{-{j_k}/Q_k} \omega_{0}^{\langle k \rangle}$. In other words, the frequency axis is divided in a (base-two) logarithmic manner. However, for $Q > 1$ in order to cover the entire frequency spectrum the first $J$ filters (i.e., $\psi_0, \psi_1, \ldots, \psi_{J-1}$) cover the higher-frequency interval in a logarithmic manner, and the lower-frequency interval is covered by $P$ equally-spaced filters with the same bandwidth as $\psi_{J-1}$. This is due to the fact that the filter $\psi_{J-1}$ has the smallest bandwidth in frequency and the largest time-support which should be smaller than the predefined 5-second translation invariant scale. Although these $P$ filters are not dilations of $\psi_{J-1}$, for simplicity they are still called wavelets~\cite{anden2011multiscale}. In this work for the first filter bank, $J=13$ and $P=1$, and for the second filter bank $J=8$ and since $Q=1$, $P=0$ (see Fig.~\ref{fig:filterbanks}). Derived by zooming in the $0-5$~Hz frequency interval of the first filter bank, Fig.~\ref{fig:filterbanks2} shows that $\psi_{13}$ has the same bandwidth as $\psi_{12}$, but the bandwidth of the other filters increases exponentially. The time-domain representations of the complex wavelets corresponding to the analytical filters in Fig.~\ref{fig:filterbanks2} are demonstrated in Fig.~\ref{fig:filterbanks3}.

The 2-layer scattering network used in this work can be summarized as follows 
\begin{eqnarray}
 S_0 x &= & x \star \phi \label{S0}\\ 
        U_1 x &= &  |x \star \psi_{j_1}| \label{U1}\\
         S_1 x &= &  |x \star \psi_{j_1}|\star \phi \label{S1}\\ 
         U_2 x &= &  ||x \star \psi_{j_1}|\star \psi_{j_2}| \label{U2}\\
         S_2 x &= &  ||x \star \psi_{j_1}|\star \psi_{j_2}|\star \phi \label{S2},
\end{eqnarray}
where $\star$ is convolution and $|\cdot|$ is the complex modulus operator. In (\ref{S0}) the zeroth-order scattering coefficient is calculated by low-pass filtering (or weighted time-averaging) of the original signal $x$ (i.e., by convolution of $x$ with the approximation function $\phi$). By this low-pass filtering, high-frequency content of $x$ is lost. This high-frequency content can be recovered by the wavelet transform. So, in (\ref{U1}) the variation of signal $x$ at different $j_1$ scales is calculated by convolution of $x$ with wavelets $\psi_{j_1}$. At a first glance it seems that the complex modulus operator $|\cdot|$ in (\ref{U1}) results in information loss as well, but it can be shown that at least for a specific family of wavelets, $x$ can be reconstructed from $|x \star \psi_{j}|$ up to a global phase (i.e., up to multiplication by a unitary complex number) and the reconstruction operator is continuous (but not uniformly continuous)~\cite{mallat2015phase}. So, the main source of information loss is the low-pass filtering which is needed for generating shift-invariant features. In (\ref{S1}) the first-order scattering coefficients are calculated by low-pass filtering of the first-order wavelet scattering modulus $U_1x$, and yet again the lost information is recovered in (\ref{U2}) in which the next wavelet scattering modulus is calculated by convolution of $U_1x$ with the second layer wavelets $\psi_{j_2}$. Finally, in (\ref{S2}) the second-order scattering coefficients are calculated. This process can be repeated an arbitrary number of times to generate more and more shift-invariant features. However, we stop it after generating the second-order scattering coefficients since the higher order coefficients have very low energy which can be neglected in the analysis~\cite{waldspurger2017exponential}, and they do not contribute towards improving the classification results~\cite{anden2014deep}. This structure mimics a CNN in the sense that the  convolutional layers (i.e., wavelet transforms $x \star \psi$) are followed by nonlinearities (i.e., modulus operations $|\cdot|$), and then they are followed by average pooling (i.e., low-pass filtering $|\cdot|\star \phi$). However, it is different from a CNN mainly because filters are not data-driven but predefined, and there is no weight sharing among different scales. 

In this work, we extract wavelet scattering coefficients for 6 biosignals: EOG, abdominal, chest, airflow, SaO2, and ECG (PSG channels 7, 9, 10, 11, 12, and 13). Since non-orthogonal Gabor wavelets have significant overlap in the frequency domain (see Figs.~\ref{fig:filterbanks} and~\ref{fig:filterbanks2}), the resulting scattering coefficients are redundant. In order to speed up the analysis and decrease the memory usage, we downsample the features by a factor of 4. The final number of features for each biosignal is 65, resulting in a total number of 390 features. 

The last point to discuss here is that one should not misinterpret our so-called physiology agnostic feature extraction as a domain agnostic method. We use the term ``physiology agnostic'' to highlight the fact that these features are not inspired by physiological knowledge of biosignals. However, the scattering transform is not a true domain agnostic method since the discovery of invariants and stability conditions to deformations which has a pivotal role in the success of this transformation is domain-dependent. It is obvious that invariants and stability conditions for image and texture data such as spatial translation, rotation, scaling, and partial occlusion~\cite{bruna2013invariant,sifre2013rotation} are different for audio and speech signals such as time shifting, time warping deformation, frequency transposition, and frequency warping~\cite{anden2014deep,anden2018classification}.

\section{Classification}\label{sec:classification}

\begin{figure}[t!]
\centering

\begin{tikzpicture}[
    font=\sf \scriptsize,
    >=LaTeX,
    cell/.style={
        rectangle, 
        rounded corners=5mm, 
        draw,
        very thick,
        },
    operator/.style={
        circle,
        draw,
        inner sep=-0.5pt,
        minimum height =.2cm,
        },
    function/.style={
        ellipse,
        draw,
        inner sep=1pt
        },
    ct/.style={
        line width = .75pt,
        minimum width=1cm,
        inner sep=1pt,
        },
    gt/.style={
        rectangle,
        draw,
        minimum width=4mm,
        minimum height=3mm,
        inner sep=1pt
        },
    mylabel/.style={
        font=\scriptsize\sffamily
        },
    ArrowC1/.style={
        rounded corners=.0cm,
        thick,
        },
    ArrowC2/.style={
        rounded corners=.0cm,
        thick,
        },
    ]

    \node [cell, minimum height =4cm, minimum width=6cm] at (0,0){} ;

    \node [gt] (ibox1) at (-2.2,-0.75) {$\sigma(\cdot)$} coordinate (iii2);
    \node [gt] (ibox2) at (-1.5,-0.75) {$\sigma(\cdot)$};
    \node [gt, minimum width=1cm] (ibox3) at (-0.5,-0.75) {tanh$(\cdot)$};
    \node [gt] (ibox4) at (0.8,-0.75) {$\sigma(\cdot)$};

\node  (gGate) at (-0.25,-0.1) {$\boldsymbol{g}_{t}$};
\node  (oGate) at (1,0) {$\boldsymbol{o}_{t}$};
\node  (iGate) at (-1.2,0.6) {$\boldsymbol{i}_{t}$};
\node  (fGate) at (-2,0.6) {$\boldsymbol{f}_{t}$};

    \node [operator] (mux1) at (-2.2,1.5) {$\otimes$};
    \node [operator] (add1) at (-0.5,1.5) {+};
    \node [operator] (mux2) at (-0.5,0.4) {$\otimes$};
    \node [operator] (mux3) at (1.5,-0.2) {$\otimes$};
    \node [function] (func1) at (1.5,0.65) {tanh$(\cdot)$};

    \node[ct, label={[mylabel]}] (c) at (-4,1.5) {$\boldsymbol{c}_{t-1}$};
    \node[ct, label={[mylabel]}] (h) at (-4,-1.5) {$\boldsymbol{h}_{t-1}$};
   \node[ct, label={[mylabel]}] (x) at (-2.1,-3) {$\boldsymbol{x}_t$};

    \node[ct, label={[mylabel]}] (c2) at (4,1.5) {$\boldsymbol{c}_t$};
    \node[ct, label={[mylabel]}] (h2) at (4,-1.5) {$\boldsymbol{h}_t$};
    \node[ct, label={[mylabel]}] (x2) at (2.5,3) {$\boldsymbol{h}_t$};
    \node(n1) at (1.5,1.5) {$\bullet$};
    \node(n2) at (2.5,-1.5) {$\bullet$};

    \draw [->, ArrowC1] (c) -- (mux1) -- (add1) -- (c2);

    \node(n31) at (-2.1,-1.7) {$\bullet$};
    \draw [ArrowC1] (-2.1,-1.7) -- (n31|-ibox1.south);
    \node(n32) at (-1.4,-1.7) {$\bullet$};
    \draw [ArrowC1] (-1.4,-1.7) -- (n32|-ibox2.south);
    \node(n33) at (-0.4,-1.7) {$\bullet$};
    \draw [ArrowC1] (-0.4,-1.7) -- (n33|-ibox1.south);
    \node(n34) at (0.9,-1.7) {};
    \draw [ArrowC1] (0.9,-1.7) -- (n34|-ibox2.south);
    
    \draw [ArrowC1] (x) |- (0.9,-1.7);

     \node(n41) at (-2.3,-1.5) {$\bullet$};
    \draw [ArrowC1] (-2.3,-1.5) -- (n41|-ibox1.south);
    \node(n42) at (-1.6,-1.5) {$\bullet$};
    \draw [ArrowC1] (-1.6,-1.5) -- (n42|-ibox2.south);
    \node(n43) at (-0.6,-1.5) {$\bullet$};
    \draw [ArrowC1] (-0.6,-1.5) -- (n43|-ibox1.south);
    \node(n44) at (0.7,-1.5) {};
    \draw [ArrowC1] (0.7,-1.5) -- (n44|-ibox2.south);
    
    \draw [ArrowC1] (h) |- (0.7,-1.5);
    
    \draw [->, ArrowC2] (ibox1) -- (mux1);
    \draw [->, ArrowC2] (ibox2) |- (mux2);
    \draw [->, ArrowC2] (ibox3)  -- (mux2) ;
    \draw [->, ArrowC2] (ibox4) |- (mux3);
    \draw [->, ArrowC2] (mux2) -- (add1);
    \draw [->, ArrowC1] (add1 -| func1)++(-0.5,0) -| (func1);
    \draw [->, ArrowC2] (func1) -- (mux3);

   \draw [->, ArrowC2] (mux3) |- (h2);
    \draw (c2 -| x2) ++(0,-0.1) coordinate (i1);
   
    \draw [->, ArrowC2] (h2 -| x2)++(-0.5,0) -| (x2);

\end{tikzpicture}

{\caption{LSTM memory cell with forget gate $\boldsymbol{f}_t$ as proposed in~\cite{Gers2000}. In the original LSTM, proposed in~\cite{hochreiter1997long}, there was no forget gate.}\label{fig:LSTM}}
\end{figure}
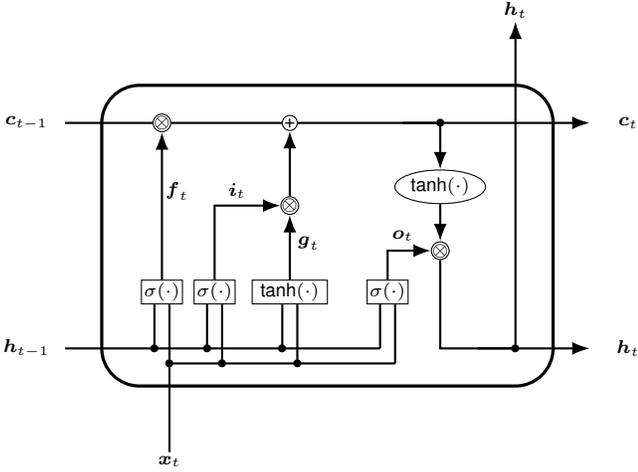

\begin{figure*}[t!]
	\centering
	\includegraphics[width=0.8\linewidth]{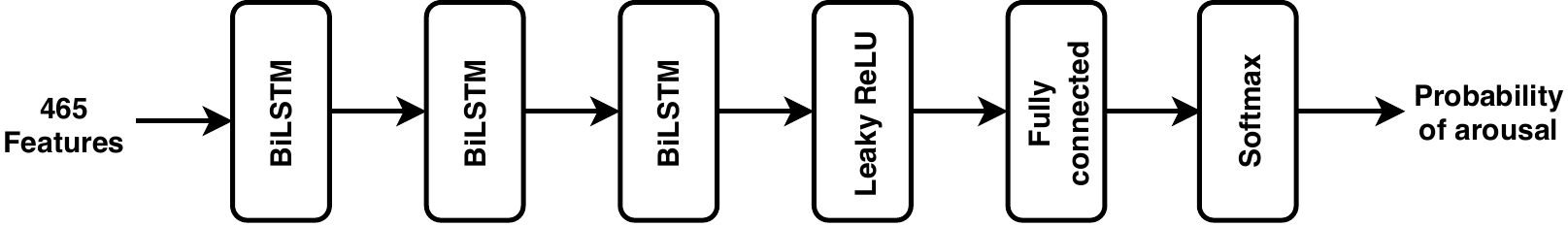}
	{\caption{The architecture of the proposed BiLSTM network.}\label{fig:BD}}
\end{figure*}

In the intermediate phase of this work after feature engineering, we relied on the sliding window method~\cite{dietterich2002machine} to classify each 5-second segment of the PSG data using a random forest classifier~\cite{breiman2001random}. On average, the best-achieved result was 0.18 in terms of AUPRC, with high variance among different chunks of the data. The main reason for this low performance is that the temporal information and dependencies among different segments of the time series are lost. To address this shortcoming we use an LSTM network~\cite{hochreiter1997long} which is a type of RNN with a gating mechanism that controls the flow of information~\cite{Goodfellow2016deep}. Contrary to ``vanilla'' RNN which suffers from the vanishing and exploding gradient problem~\cite{bengio1994learning} and consequently does not capture the long-range dependencies, LSTM addresses the aforementioned problem and captures richer contextual information of the time series, thanks to the gating mechanism.

In this work, we analyze the PSG recordings retrospectively and since the past, present, and future information of the time series is available at analysis time, we can use a bidirectional LSTM (BiLSTM) variant. Each BiLSTM layer consists of two layers of LSTMs: causal and anticausal counterparts. A single unit of a causal LSTM which processes the time series forward in time is illustrated in Fig.~\ref{fig:LSTM}. It consists of four gates that control the flow of information through the following equations 
\begin{eqnarray}
         \boldsymbol{i}_t &= & \sigma \left(\boldsymbol{W}_i\boldsymbol{x}_t+\boldsymbol{U}_i\boldsymbol{h}_{t-1}+\boldsymbol{b}_i\right) \label{inputG}\\
          \boldsymbol{f}_t &= &  \sigma \left(\boldsymbol{W}_f\boldsymbol{x}_t+\boldsymbol{U}_f\boldsymbol{h}_{t-1}+\boldsymbol{b}_f\right) \label{forgetG}\\ 
          \boldsymbol{g}_t &= &  \tanh \left(\boldsymbol{W}_g\boldsymbol{x}_t+\boldsymbol{U}_g\boldsymbol{h}_{t-1}+\boldsymbol{b}_g\right) \label{candidateG}\\
          \boldsymbol{o}_t &= &  \sigma \left(\boldsymbol{W}_o\boldsymbol{x}_t+\boldsymbol{U}_o\boldsymbol{h}_{t-1}+\boldsymbol{b}_o\right) \label{outputG}\\
          \boldsymbol{c}_t &= &  \boldsymbol{f}_t \otimes \boldsymbol{c}_{t-1} + \boldsymbol{i}_t \otimes \boldsymbol{g}_t\label{cellState}\\
          \boldsymbol{h}_t &= &  \boldsymbol{o}_t \otimes \tanh(\boldsymbol{c}_t)\label{hiddenState}.
\end{eqnarray}
Here, $\boldsymbol{i}$, $\boldsymbol{f}$, $\boldsymbol{g}$, and $\boldsymbol{o}$ are the vectors related to the input gate, forget gate, candidate cell gate, and output gate, respectively, for the entire layer of units or memory cells. Moreover, vectors $\boldsymbol{c}$ and $\boldsymbol{h}$ are the cell and hidden states, respectively, and $\boldsymbol{W}_*$, $\boldsymbol{U}_*$, and $\boldsymbol{b}_*$ are respectively the input weight matrix, recurrent weight matrix, and bias vector for the gate denoted by $* \in \{i,f,g,o\}$. $\sigma(\cdot)$ and $\tanh(\cdot)$ denote respectively sigmoid and hyperbolic tangent activation functions, and $\otimes$ is the Hadamard product. The anticausal LSTM which processes the time series backward in time is similar to the forward LSTM with reverse time order which leads to similar equations with different weights and biases ($\boldsymbol{W'}_*$, $\boldsymbol{U'}_*$, $\boldsymbol{b'}_*$); moreover, $\boldsymbol{h}_{t-1}$ and $\boldsymbol{c}_{t-1}$ are replaced respectively by $\boldsymbol{h'}_{t+1}$ and $\boldsymbol{c'}_{t+1}$. The outputs of the two LSTMs are then concatenated to capture the contextual information of the whole time series.

The architecture of the proposed BiLSTM network is illustrated in Fig.~\ref{fig:BD} in which 3 layers of BiLSTMs with 400 hidden units per layer (200 for each LSTM) are followed by a leaky rectifier linear unit (Leaky ReLu) layer, a fully connected layer, and a softmax layer. We have scrutinized and evaluated several different combinations, to empirically identify the best architecture. To name a few, we have examined a different number of BiLSTM layers, different number of memory cells per layer, multiple activation functions (e.g., linear, ReLu, Leaky ReLu, and sigmoid), different number of fully connected layers, and different parameters for Leaky ReLu layer. Leaky Relu has the following equation 
\begin{equation}
\label{cases}
f(x)=\begin{cases}
x  & \text{for } x> 0,\\
 a x & \text{for } x\leq 0,
 \end{cases}
\end{equation}
where typically $a$ is a small number (e.g., 0.01). However, we have obtained the best result with $a = 0.5$. The theoretical reason behind this observation is not clear which is not an uncommon situation in the field of deep learning. Although there are studies which discuss the effect of different nonlinearities~\cite{djork2016fast,he2015delving,xu2015empirical}, they mainly focus on CNNs and suffer from the lack of mathematical rigor.

We have also applied the dropout mechanism~\cite{hinton2012improving,srivastava2014dropout} between different layers of the network, but the classification accuracy declined. We implemented our proposed method in MATLAB R2018b which only has an input/output dropout layer. However, for RNNs there is a more effective type of dropout mechanism which is applied to recurrent layers~\cite{gal2016theoretically}. In fact, since the employed dropout mechanism was not useful, we decided not to use it and instead selected a set of discriminative features before feeding them to the BiLSTM network. For physiology informed features we proposed a heuristic feature selection method as follows. At first, we pre-train the BiLSTM network with 500 features, and rank them using the following ad hoc score:
\begin{equation}
S_k =  \sum_{*\in\{i,f,g,o\}} \sum_{j=1}^N \Big(|W_*(j,k)|+|W'_*(j,k)|\Big)\label{FS_network}, 
\end{equation}
and then select the 75 top-ranked features. In~(\ref{FS_network}), $W_*(j,k)$ and $W'_*(j,k)$ are the weights of the connections between the $k$-th feature and the $j$-th memory cell of the forward and backward LSTMs in the first BiLSTM layer, respectively, $|\cdot|$ is the absolute value, and $N=200$ is the number of memory cells of each LSTM. For physiology agnostic features we do not apply any feature selection and feed them (390 features) directly to the network.

\begin{table*}[ht!]
\begin{center}
\caption{The Classification Performance of All Features (submitted for PhysioNet challenge) together with Physiology Informed and Physiology Agnostic Features}\label{table:DetailResults}

\begin{tabular}{@{}c c c c  c c c  }
\toprule[0.12em]
  	 {}& \multicolumn{2}{c}{All features} & \multicolumn{2}{c}{Physiology informed} &\multicolumn{2}{c}{Physiology agnostic} \\
  	 {}& \multicolumn{2}{c}{(n = 465)} & \multicolumn{2}{c}{features (n = 75)} &\multicolumn{2}{c}{features (n = 390)} \\
\cmidrule[0.06em](lr){2-3} \cmidrule[0.06em](lr){4-5} \cmidrule[0.06em](lr){6-7} 
    {Training data} &  AUPRC    & AUROC  & AUPRC   & AUROC  & AUPRC   & AUROC  \\ %\hline
	 \cmidrule[0.06em]{1-1} \cmidrule[0.06em](lr){2-3} \cmidrule[0.06em](lr){4-5} \cmidrule[0.06em](lr){6-7} 
fold 1~   &   0.46   &  0.90  & 0.50   &   0.92  &   0.42   &   0.89      \\  
	
fold 2~   &   0.48   &  0.89  & 0.47   &   0.92  &   0.53   &   0.90      \\ 
	
fold 3~   &   0.50   &  0.91  & 0.56   &   0.93  &   0.49   &   0.89      \\ 
	
fold 4~   &   0.48   &  0.90  & 0.55   &   0.92  &   0.46   &   0.89      \\
	
fold 5~   &   0.43   &  0.91  & 0.51   &   0.92  &   0.46   &   0.88      \\ 
	
fold 6~	 &   0.55   &  0.91  & 0.53   &   0.91  &   0.50   &   0.90     \\
	
fold 7~   &   0.54   &  0.90  & 0.51   &   0.92  &   0.43   &   0.90      \\

fold 8~   &   0.52   &  0.91  & 0.44   &   0.90  &   0.45   &   0.88      \\
	
fold 9~   &   0.52   &  0.91  & 0.49   &   0.91  &   0.44   &   0.90      \\

fold 10  &   0.42   &  0.89  & 0.50   &   0.92  &   0.40   &   0.87      \\  
	
	\cmidrule[0.06em]{1-1} \cmidrule[0.06em](lr){2-3} \cmidrule[0.06em](lr){4-5} \cmidrule[0.06em](lr){6-7} 
	
Mean  & 0.49 & 0.90& \textbf{0.51}   &  \textbf{0.92}  & 0.46  & 0.89 \\ 

(STD)  & (0.04) & (0.01)& (0.04)   &  (0.01)   & (0.04)  & (0.01)  \\

\cmidrule[0.06em]{1-1} \cmidrule[0.06em](lr){2-3} \cmidrule[0.06em](lr){4-5}  \cmidrule[0.06em](lr){6-7} 
\textbf{Test data} & \textbf{0.50} & --- & ---   &  ---   & ---  & ---  \\ 
 	\bottomrule[0.12em]
	\end{tabular}
	\end{center}
\end{table*}

Before feeding the network with training data, all PSG segments with non-target arousal labels are removed. Then, the recordings were sorted based on the feature sequence length. The sorted data are further divided into mini-batches with a size of 20 subjects. Feature sequences inside each mini-batch are zero-padded in order to have the same length. The network is trained by these mini-batches to obtain the weights and biases which minimize the cross-entropy loss function using the Adam optimization algorithm~\cite{kingma2014adam}. In order to address the class imbalance problem, we use a weighted cross-entropy loss function with 0.9 and 0.1 weights for target arousal and non-arousal classes, respectively. Moreover, we use 0.005 learning rate which is 5 times larger than the default value of the Adam algorithm. By choosing this value, we obtain a better result and have a faster training phase. Other important parameters such as exponential decay rates of the first and second moment estimates are set to their default values: 0.9 and 0.999. The training is done during 30 epochs, but after every 10 epochs, the learning rate drops to 70\% of its previous value. Finally, we also employ the gradient norm clipping techniques~\cite{pascanu2013difficulty} by putting a further constraint on the gradient norm $\|\boldsymbol{g}\|$ not to be larger than 1. If $\|\boldsymbol{g}\|>1$, the gradient $\boldsymbol{g}$ is replaced by $\boldsymbol{g}/\|\boldsymbol{g}\|$. The reason for using this technique is that if the gradient has a very large value, then the update term in the gradient descent-based algorithm may cause the parameters to jump to a point far from their current position, increasing the value of the loss function, thus wasting most of the efforts made so far to reach the current point~\cite{Goodfellow2016deep}. To prevent this issue we move a smaller distance in the gradient direction.

\section{Results}\label{sec:results}

The performance of our proposed method, consisting of 465 features and a BiLSTM network, for classification of PSG data for sleep arousal detection is assessed on the training dataset using a 10-fold cross-validation procedure. The proposed method achieves average scores of 0.49 and 0.90 for AUPRC and AUROC, respectively. Moreover, to evaluate the performance on the test dataset with hidden labels, the ensemble of the BiLSTM networks, trained on the aforementioned 10-fold cross-validation committee, is submitted to the PhysioNet system. The ensemble classifier achieves the state-of-the-art AUPRC score of 0.50, which is the second-best score during the follow-up and official phases of the 2018 PhysioNet challenge. This result is also 0.31 points better than our prior work~\cite{zabihi2018automatic}.

Table~\ref{table:DetailResults} shows the classification performance of different sets of features using the same BiLSTM architecture. The set of physiology informed features achieves the best average scores of 0.51 AUPRC and 0.92 AUROC, even better than our submitted method. This result is the same as the result of the winner algorithm of the 2018 PhysioNet challenge~\cite{howe2018automated} on the training dataset. The performance of the physiology agnostic features (i.e., scattering transform features) is worse than the results of the physiology informed features by 0.05 and 0.03 points in terms of AUPRC and AUROC, respectively. However, it is still among the top 5 PhysioNet algorithms on the training dataset.

\begin{table*}[t!]
\begin{center}
\caption{The Classification Results of Different Sets of Features on the Training Dataset}\label{table:FeatureResults}
\footnotesize
\begin{tabular}{l c c c  c   }
\toprule[0.12em]
  	 {Feature type} &     Number of features & \# PSG channel  & AUPRC   & AUROC    \\ \hline

	 1)	cross-channel    &   13   &  9-11  & 0.42 (0.04)   &   0.88 (0.01)     \\
	 2)	abdominal  &   6   &  9  & 0.38 (0.04) &   0.86 (0.01)     \\
	 	 3)	chest  &   5   &  10  &      0.29 (0.03)  & 0.82 (0.01)  \\
	 4) airflow &            12   &  11  & 0.29 (0.03)   &   0.79 (0.02)  \\

	 5)   SaO$_2$   &   5   &  12  & 0.28 (0.05)   &   0.82 (0.04)      \\

	 	6)  	EEGs    &   22   &  1-6  & 0.28 (0.04)   &   0.82 (0.01)     \\  
	 	7) 	EOG    &   7   &  7  & 0.22 (0.04)   &   0.77 (0.02)     \\  
	 	 8) chin EMG + ECG    &   5   &  8,13  & 0.25 (0.03)  &   0.80 (0.01)     \\

	    \cmidrule[0.06em]{1-1} \cmidrule[0.06em](lr){2-2} \cmidrule[0.06em](lr){3-3}  \cmidrule[0.06em](lr){4-4}  \cmidrule[0.06em]{5-5} 
	    
	 Respiratory-related    &   41   &  9-12  & 0.46 (0.04)   &   0.90 (0.01)       \\ 
	 
	 Non-respiratory-related   &   34   &  1-8,13  & 0.33 (0.03)   &   0.85 (0.01)      \\ 
	 
	    \cmidrule[0.06em]{1-1} \cmidrule[0.06em](lr){2-2} \cmidrule[0.06em](lr){3-3}  \cmidrule[0.06em](lr){4-4}  \cmidrule[0.06em]{5-5} 
	 Physiology informed     &   75   &  1-13  & \textbf{0.51 (0.04)}   &   \textbf{0.92 (0.01)}     \\  
	 Physiology agnostic     &   390   &  7,9-13  & 0.46 (0.04)   &   0.89 (0.01)     \\

	\cmidrule[0.06em](lr){1-1} \cmidrule[0.06em](lr){2-2} \cmidrule[0.06em](lr){3-3} \cmidrule[0.06em](lr){4-4}  \cmidrule[0.06em]{5-5}

	 	All     &   465   &  1-13  & 0.49 (0.04)   &   0.90 (0.01)     \\

 	\bottomrule[0.12em]
	\end{tabular}

\end{center}

\end{table*}

The performance of the physiology informed features may raise a question concerning our submitted method. The reason that the method with the inferior result (0.49 vs. 0.51; see Table~\ref{table:DetailResults}) is submitted for evaluation on the test dataset is that in the intermediate stage of this work in order to speed up the experiments, the performances of different methods were assessed by holdout validation strategy and the proposed method with 465 features achieved the best results. However, when we evaluate the models using the 10-fold cross-validation assessment technique we notice that the 75 physiology informed features outperform our proposed method by 0.02 points in terms of the AUPRC score. Since we only had one submission for the proposed algorithm, we could not evaluate the performance of our physiology informed features on the test dataset.

Table~\ref{table:FeatureResults} shows the detailed performance of different types of features for sleep arousal detection using 10-fold cross-validation on the training dataset. Although limited in scope, for the sake of simplicity we use the same BiLSTM network architecture for all experiments. It is clear that for a more reliable comparison, the network architecture and parameters need to be optimized for each set of features. The only parameter which is altered for different sets of features is the learning rate of the Adam optimization algorithm. 

The last points to discuss are two technicalities. First, the time resolution for analysis of the PSG data is 5 seconds. In other words, for each 5-second window, our classification algorithm generates only one label (probability of arousal) and in order to have the sample-wise probability of arousals as demanded by PhysioNet, we repeat the value 1000 ($=5\times 200$) times. Second, in Tables~\ref{table:DetailResults} and~\ref{table:FeatureResults} for training different folds whenever the optimization algorithm gets stuck at a local minimum or much more probably at a saddle point~\cite{dauphin2014identifying,choromanska2015loss} we rerun the training phase with different network initialization.

\section{Discussion}

In this study, we investigate a comprehensive set of hand-engineered features for retrospective analysis of PSG data using a BiLSTM classifier for non-apneic/non-hypopneic arousal detection. We extract multi-domain features from different modalities. During multiple steps of feature selection techniques and expert judgment, the irrelevant and/or redundant features are eliminated to obtain a set of 75 physiology informed features. The final set of 465 features are built upon these 75 and an additional set of 390 features derived using a state-of-the-art scattering transform. The features are then fed into a BiLSTM network to classify the PSG data. Our proposed method achieves the second best score of 0.50 AUPRC on the hidden test dataset of the 2018 PhysioNet challenge. In this section, we scrutinize the results and discuss ways to further improve them.

\subsection{Comparative Evaluations of Selected Features}

The best single type of features in Table~\ref{table:FeatureResults} are the cross-channel features, which achieve average scores of 0.42 and 0.88 in terms of AUPRC and AUROC, respectively. To the best of our knowledge, this is the first time that p-values and SVD-based features are proposed for analysis of respiratory effort signals (chest and abdominal) alongside respiratory airflow. The next best single type of features are the ones extracted from the abdominal-only signal with an average score of 0.38 AUPRC. The features extracted from the chest, airflow, SaO$_2$, and EEG signals have nearly 0.29 AUPRC. EOG and chin EMG have also the same performance of 0.22 AUPRC. However, since the number of features extracted from chin EMG and ECG is low, they are combined together, resulting in 0.25 AUPRC.  

The respiratory-related features, obtained by combining feature types 1-5, have a high AUPRC score of 0.46 for arousal detection. This is not surprising, considering that most of the arousals are RERAs and for detecting them respiratory-related biosignals such as airflow, chest, and abdominal play a pivotal role. However, the interesting observation is that the performance of SaO$_2$ is as good as EEG, although the degree of oxygen desaturation is not a requirement for RERA detection~\cite{Bornemann2006}. If non-respiratory-related features, obtained by combining feature types  6-8, with 0.34 AUPRC score are added to the aforementioned respiratory-related features the resulting physiology informed features have average scores of 0.51 and 0.92 in terms of AUPRC and AUROC, respectively. This is the best-achieved result among all combination of features. 

Regarding the selected EEG-based features, although AASM guidelines determine arousals as abrupt EEG frequency shifts toward rhythms such as alpha, theta, and/or beta above 16~Hz~\cite{berry2018aasm}, our experiments show that only delta (0.1-4~Hz) power is chosen as a discriminative feature for arousal detection (see Section~\ref{sec:NRRF}). This is an intriguing observation, not expected a priori. However, some studies support the hypothesis of a continuum in arousal activities which start from delta and K-complex bursts toward EEG arousals and full awakening~\cite{halasz2004nature,sforza2000cardiac}. More specifically, an increase in delta power can be a pre-arousal activity which may or may not culminate to an EEG arousal~\cite{halasz2004nature}. Furthermore, in~\cite{de2004quantitative} and~\cite{terzano2002cap} the association between arousals and K-complexes or delta bursts preceding the events are confirmed. This occurs for arousals in NREM sleep stage but not during REM. Besides, in both upper airway resistance syndrome (UARS) and obstructive sleep apnea syndrome (OSAS), airway opening is associated with an increase in delta power which can be followed by an EEG arousal~\cite{black2000upper,poyares2002arousal,berry1998within}. Since RERAs are increased in both UARS and OSAS, it might be a reason that delta power is one of the selected features, especially because we use a BiLSTM classifier capable of analyzing the sequence of transient events. However, we stress that at this point we cannot identify the causes of this observation with certainty and it requires further investigation. On top of that, we do not claim that alpha and beta powers are not important, but maybe that their information is covered by other features. For example, the RMS and STD of the signal amplitude can partially retain alpha and beta information. Recall that alpha and beta are low-amplitude high-frequency EEG rhythms.  

Another interesting observation is that contrary to the AASM guidelines which recommend central and occipital EEG channels as primary signals for detecting EEG arousals, in our experiments most of the EEG-based features are selected from frontal channels. This might be partially advocated by the fact that delta activity and K-complexes predominate in the frontal lobe of the brain~\cite{McCormick1997,accolla2011clinical,maurer2012atlas}. In addition, in~\cite{younes2016accuracy} the authors show that for in-home sleep stage scoring the analysis of either frontal or central EEG channels leads to similar results when the recordings are scored by automatic Michele Sleep Scoring (MSS) system~\cite{malhotra2013performance,younes2015utility}.

Table~\ref{table:FeatureResults} also shows that the performance of the physiology agnostic features based on the scattering transform is less than the performance of the physiology informed features by 0.05 and 0.03 points in terms of AUPRC and AUROC, respectively. This may have the following possible causes. First, in order to decrease the computational complexity in calculating the scattering transform, only 6 PSG time series are used (i.e., the remaining 7 time series including 6 EEGs and 1 chin EMG are not used). Second, to expedite the analysis and decrease the memory usage, the extracted features are downsampled by a factor of 4. Third, we believe that the way we treat the scattering transform as a physiology agnostic method restricts the performance of this set of features by not including any prior physiological information in designing the filter banks; we use the same filter banks for all biosignals (see Fig.~\ref{fig:filterbanks}). However, we already saw in Sections~\ref{PIF} that different clinical time series carry important information in different frequency bands. Forth, as we discussed earlier in Section~\ref{sec:ST} the invariants and stability conditions for different biosignals need to be explored for achieving the optimal performance of the scattering transform. 

Although our first motivation to use the scattering transform was to semi-automatically derive a set of informative features with minimum expert intervention, a more efficient approach is to include minimum physiological information at least in designing the filter banks for different biosignals. Despite the aforementioned constraint imposed by us on using the scattering transform, yet again we underline that even with this suboptimal handling of this method, the result solely based on the scattering transform is still among the top 5 algorithms on the training dataset. Future developments may include the design of the optimal filter banks for each biosignal separately.

\subsection{Adaptation for Real-Time Classification}

 Although the proposed algorithm is designed for retrospective analysis of PSG data, with minor adaptation it can be used for real-time classification, possibly with an only 5-second delay. It is worth mentioning that all top-ranked algorithms for sleep arousal detection in PhysioNet 2018 challenge \cite{thrainsson2018automatic,He2018identification,varga2018using,patane2018automated} use longer analysis windows. The 5-second analysis window used by our algorithm makes it a potential candidate for real-time classification if needed. For real-time classification, we need to design the artifact removal filter based on the information in the current and/or the previous time-windows. To be more precise, we need to remove all future information which is available in the present version of our algorithm; in the current setup, the threshold for the artifact removal filter for each biosignal is calculated from the information of the entire time series (see section~~\ref{preprocessing}). After this stage, we only need to replace the BiLSTM layers by LSTM ones in the classifier (see Fig.~\ref{fig:BD}). In that case, our algorithm can classify the PSG data in real-time. However, its performance degrades drastically by 0.13 and 0.05 points to 0.36 and 0.85 in terms of AUPRC and AUROC, respectively. This is to be expected since future information of the time series is not utilized in the LSTM network and may also be due to the non-optimized network architecture and parameters for the new setup.

\subsection{Study Limitations}

The main limitation of the study is the annotation process of the PhysioNet dataset. During the labeling process, seven sleep technologists annotated the dataset. However, given the burden involved in manual annotation, each PSG recording was annotated by only one sleep technologist~\cite{ghassemi2018you}. This calls into question both the consistency and the reliability of the annotations. Although the performance of our proposed method indicates that our algorithm can replicate the experts' annotations accurately, its medical significance is limited by the labeling process. It is clear that high inter- and intra-rater agreement between sleep technologists would lead to a more reliable annotated dataset. This consequently results in greater clinical significance. Moreover, the number of submissions and the run-time constraint, imposed by the PhysioNet challenge rules, are the other limitations of this study.

\section{Conclusion}

We have designed and implemented an automated PSG-based classification algorithm to detect non-apneic and non-hypopneic arousals. We have demonstrated and validated its performance using the 2018 PhysioNet challenge dataset, which is the newest and largest publicly available PSG dataset. Our proposed algorithm has tied for the second-best score during the follow-up and official phases of the 2018 PhysioNet challenge, by achieving the state-of-the-art performance of 0.50 in terms of AUPRC. 

In this study, we have also paid special attention to extracting features based on the physiological process during RERAs, which is missing in typical end-to-end deep learning algorithms. We have investigated and evaluated the importance of different types of features for automatic arousal detection. We have had interesting findings regarding the selected features, which were not expected a priori and may contribute to a better understanding of the RERAs, helping us for developing new automated algorithms. Besides, we have developed an alternative semi-automatic PSG-based feature extraction method using scattering transform and discussed possible directions for improving the performance.

\section{Acknowledgment}

The authors would like to thank Leo K\"{a}rkk\"{a}inen for his insightful and valuable comments on this work.

\bibliographystyle{IEEEtran}
\bibliography{refs}

\end{document}